\documentstyle[12pt]{article}
\begin{document}

\newcommand\KENT{1}
\newcommand\WIRT{2}
\newcommand\WERN{3}
\newcommand\LEDE{4}
\newcommand\MALL{5}
\newcommand\RIKV{6}
\newcommand\RICHm{7}
\newcommand\RICHd{8}
\newcommand\RICHb{9}
\newcommand\RIKVr{10}
\newcommand\NOVO{11}
\newcommand\VOTE{12}
\newcommand\AHAR{13}
\newcommand\BROWWF{14}
\newcommand\BIND{15}
\newcommand\MART{16}
\newcommand\BOER{17}
\newcommand\GARA{18}
\newcommand\BROWG{19}
\newcommand\NEEL{20}

\begin{center}
{\bf 
COMPUTER SIMULATIONS OF THERMAL SWITCHING IN SMALL-GRAIN FERROMAGNETS.}\\
\end{center}

\vskip 15 truept
\noindent M.~A.\ NOVOTNY$^{*}$, G.~BROWN$^{*,**}$, 
and P.~A.\ RIKVOLD$^{*,**}$ 

\noindent *Supercomputer Computations Research Institute, 
Florida State U., 
Tallahassee, FL 32306-4130, USA; ~ novotny@scri.fsu.edu; ~ 
browngrg@scri.fsu.edu; ~ rikvold@scri.fsu.edu

\noindent **Center for Materials Research and Technology and Department of 
Physics, Florida State U., Tallahassee, FL 32306--4350

\vskip 15 truept
\noindent {\bf ABSTRACT}

We present Monte Carlo and Langevin micromagnetic calculations to
investigate thermal switching of single-domain ferromagnetic 
particles.  
For the Monte Carlo study we 
place particular emphasis on the probability that
the magnetization does not switch by time $t$, $P_{\rm not}(t)$.  
We find that $P_{\rm not}(t)$ has different behaviors in different
regimes of applied field, temperature, and system size, and we 
explain this in terms of different reversal mechanisms that dominate in the 
different regimes.  In the micromagnetic study of an array of Ni pillars, we 
show that the reversal mode is an `outside-in' mode starting 
at the perimeter of the array of pillars.  

\vskip 15 truept
\noindent {\bf INTRODUCTION}

All facets of the dynamics of nanoscale magnetic materials 
are currently active areas of research.  
The ability to construct 
single-domain nanoparticles via various methods 
and to measure the properties of individual nanoparticles 
and small arrays of nanoparticles [\KENT \ --- \LEDE ] 
provides clean experiments compared with 
previous studies of mixtures of magnetic particles.  
One of the driving forces for applications is the rapid increase in 
density of magnetic recording devices, and the associated need to 
store each bit of information on a smaller number of grains [\MALL ].  
On the theoretical and simulational side, a detailed understanding of 
nucleation and growth mechanisms that lead to the 
decay of a metastable state 
in finite systems has led to the identification of different decay 
modes in different parameter regimes [\RIKV \ --- \RIKVr ].  
In addition, new simulation algorithms are becoming available that 
may allow microscopic simulations at the inverse phonon frequency 
to extend to the technologically important time scales of years 
[\NOVO , \VOTE ].  

In this brief paper we concentrate on using the thermal activation picture of 
nucleation and growth in simple metastable systems to 
better understand the reversal mechanisms for more realistic 
models of magnetism.  We present 
Monte Carlo simulations for $P_{\rm not}(t)$, the probability that 
the metastable magnetization has not yet 
reversed at time $t$.  
We show how well this simulation of a heterogeneous system 
fits our theoretical description for $P_{\rm not}(t)$.  
We also present Langevin micromagnetic results for an array of 
Ni pillars.  
In the Langevin micromagnetic calculations 
the switching mechanism involves 
escape over a saddle point driven by random thermal fluctuations at 
constant field, 
rather than the deterministic disappearance of the metastable state 
as the field increases [\AHAR , \BROWWF ].  
Consequently, the Langevin micromagnetic calculations 
have the ability to measure $P_{\rm not}(t)$ directly.  

\vskip 15 truept
\noindent {\bf MODELS AND METHODS}

We have preformed two types of computer calculations.  
The first consists of Monte Carlo simulations [\BIND ] of the 
square-lattice Ising model 
with Hamiltonian 
\begin{equation}
{\cal H} = - J\sum_{\langle i,j\rangle} s_i s_j - \sum_i H_i s_i \;,
\end{equation}
where $s_i$$=$$\pm 1$ and the local fields $H_i$ are  
random numbers uniformly distributed between a 
maximum and a minimum value.  
This work uses periodic boundary conditions and a 
Glauber Monte Carlo update at randomly chosen sites.  
The initial state has all spins up, and at $t$$=$$0$ a negative 
field $\{H_i\}$ is applied.  
The unit of time is Monte Carlo Steps per Spin (MCSS).  
A rigorous derivation
of the stochastic Glauber dynamic for Ising models
from microscopic quantum Hamiltonians
has been established under certain conditions
in the thermodynamic limit [\MART], with 
the Monte Carlo time unit related to heat-bath phonon frequencies.  
This simulation has been performed to test our prediction for 
the forms of $P_{\rm not}(t)$ in a system with 
quenched bulk randomness.  

In order to simulate models with realistic spin degrees of freedom, 
we have programmed a Langevin micromagnetics
code similar to that reported in [\BOER ].
We have used a phenomenological damping parameter $\alpha$,
and classical spins of constant length given by the 
bulk saturation magnetization $M_{\rm s}$.  
Then at each lattice site $i$ there is a scaled magnetization
${\vec m}={\vec M}_{\rm s}/M_{\rm s}$.
The standard Ginzburg-Landau-Lifshitz micromagnetic equation 
[\AHAR , \BROWWF ] is
\begin{equation}
{{d{\vec m}_i}\over{dt}} =
-{1\over{1+\alpha^2}}
{\vec m}_i\times
\left({\vec h}_{i,{\rm eff}}+\alpha{\vec m}_i\times
{\vec h}_{i,{\rm eff}}\right)
\; .
\end{equation}
The scaled effective field at each site, ${\vec h}_{i,{\rm eff}}$,
contains contributions from terms including
the exchange interaction,
the dipole-dipole interaction,
the interaction due to crystalline anisotropy,
the applied field, and
a scaled noise term proportional to the
the Langevin fields $\zeta(t)$ [\BROWWF , \BOER ].
The Langevin noise term $\zeta$ and
the integration time step $\Delta t$
are related by $\zeta$$\propto$$\sqrt{\Delta t}$.  
Even though the set of equations used in this Langevin micromagnetics
simulation are approximations to the actual equations [\GARA ],
the approximation should be reasonable far below
the critical temperature.  
We have used a fourth-order Runge-Kutta algorithm 
as the integration scheme in 
order to keep the length of the ${\vec m}_i$ constant.  

\vskip 15 truept
\noindent {\bf RESULTS}

\begin{figure}[t]
\vspace{7.5truecm}
\includegraphics{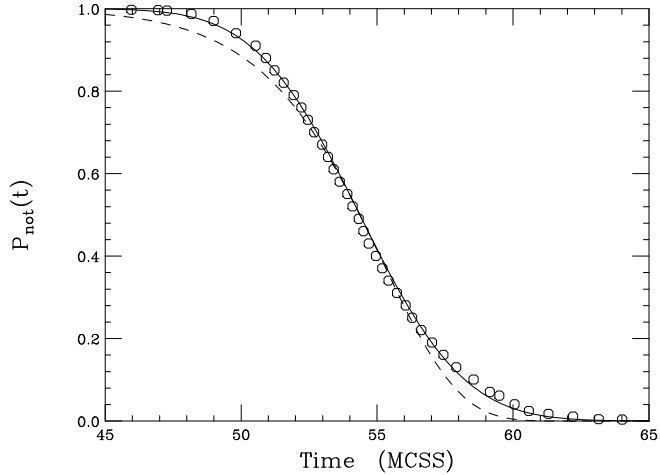}
\caption[]{
The probability $P_{\rm not}(t)$ in the multidroplet (MD) regime 
for a $100$$\times$$100$ Ising model 
with periodic boundary conditions at $T$$=$$0.8$$T_{\rm c}$.  
The data (open circles) are for 1000 escapes from the metastable state.  
The random fields $H_i$ are uniformly distributed, and centered on 
$-0.34725$ with a width of $0.34725$.  
Two different one-parameter fits to the data are shown.  The solid 
line 
is a fit to the complementary error function given by Eq.~(3).  
The dashed curve is a one-parameter fit to a stretched exponential.  
In both of these fits the average lifetime 
was set at the measured value of $\tau$$=$$54.4$~MCSS.  
(So that the solid line can be seen, only a small number of the 
1000 data points are shown.) 
}
\end{figure}

In the Ising simulation, we used $100$$\times$$100$ 
lattices at $T$$=$$0.8T_{\rm c}$$\approx$$1.815J$.  
As in the case of homogeneous nucleation [\RICHm ], 
homogeneous nucleation for single-domain Ising particles 
with demagnetizing fields [\RICHd ], and 
for single-domain Ising particles with different 
boundary conditions [\RICHb ], we have identified 
different decay regimes [\RIKV \ --- \RIKVr ].  
The functional form for quantities such as $P_{\rm not}(t)$ 
are different in the different decay regimes.  
In particular, in the single-droplet (SD) regime of a single-domain 
magnetic particle, 
$P_{\rm not}(t)=\exp\left(-t/\tau\right)$.  Here $\tau$ is the 
average lifetime for the decay of the metastable magnetic 
state due to thermal fluctuations.  
However, in the multi-droplet (MD) regime for 
single-domain particles the functional form for $P_{\rm not}(t)$
is given by 
\begin{equation}
P_{\rm not}(t) = {1\over 2} 
{\rm erfc}\left({{t-\tau}\over{\Delta}}\right) \; ,
\end{equation}
where ${\rm erfc}$ is the complementary error function, and 
the width $\Delta$ depends on the system size. 
This form for $P_{\rm not}(t)$ results from the assumption that 
many independent droplets nucleate and grow 
in different parts of the system,  
collectively leading to the magnetization reversal.  
Figure~1 shows data obtained from 1000 reversals with the 
fields $H_i$ 
uniformly distributed with width $0.34725$~$J$ centered about 
$-0.34725$~$J$.  This distribution for $H_i$ ensures that 
the system is in the multi-droplet regime [\RICHm ].  
This figure should be compared with Fig.~3c of 
Ref.~[\RICHm ] which is for the Metropolis dynamic with 
a uniform field.  
The lifetime $\tau$ was measured to be 
$\tau$$=$$54.4$~MCSS, where the time 
unit is Monte Carlo Steps per Spin (MCSS).  
Using the Mathematica nonlinear fit function 
to fit Eq.~(3) to the data, gives the 
value $\Delta$$=$$4.26$~MCSS, which is shown as 
a solid curve in Fig.~1.  It is also possible to 
try other standard expressions for the sigmoidal curve 
for $P_{\rm not}(t)$ in the MD regime.  Fig.~1 also 
shows a one-parameter fit to a stretched 
exponential, which gives 
$P_{\rm not}(t)=\exp\left(-b t^{20.7}\right)$, 
shown as the dashed line.  
Here $b$ is determined by requiring that the average of 
the distribution be $\tau$.  
Clearly among these one-parameter fits, the one to 
Eq.~(3) fits the data much better than does a 
stretched exponential.  

Figure~2 shows an example of 
the type of thermal switching simulations [\BROWG ] 
that can be performed using Langevin micromagnetic calculations.  
This figure represents a square array of magnetic Ni pillars.  
Similar arrays of Fe pillars have been 
built and measured experimentally [\KENT , \WIRT ].
The simulation was started with all spins 
pointing up, and at $t$$=$$0$ a field parallel to the 
spins was applied for 1~nsec to allow the system to 
come to thermal equilibrium.  
Then the field was reversed to point opposite to 
the average magnetization, leaving the spins in a
metastable state. 
However, the magnitude of the applied field would not have been 
sufficient to switch the system at zero temperature, and the 
switching event depicted in Fig.~2 is enabled by the thermal 
fluctuations included in Eq.~(2). 
The random thermal field has its strongest effect at 
the edges of the array, where the demagnetizing field is weakest.  
In particular, an `outside-in' switching mode is seen, in which 
the decay towards the stable magnetizaton direction
starts from the pillars at the edge 
and subsequently propagates to the pillars in the interior of the array. 

It is important to note that both the Monte Carlo and the 
Langevin micromagnetic simulations were conducted for applied fields 
sufficiently weak 
that a free-energy barrier against decay of the metastable state 
remained, and that this barrier had to be
overcome by the thermal fluctuations in order for the magnetization switching 
to occur.  

\begin{figure}[t]
\vspace*{13.0truecm}
\includegraphics{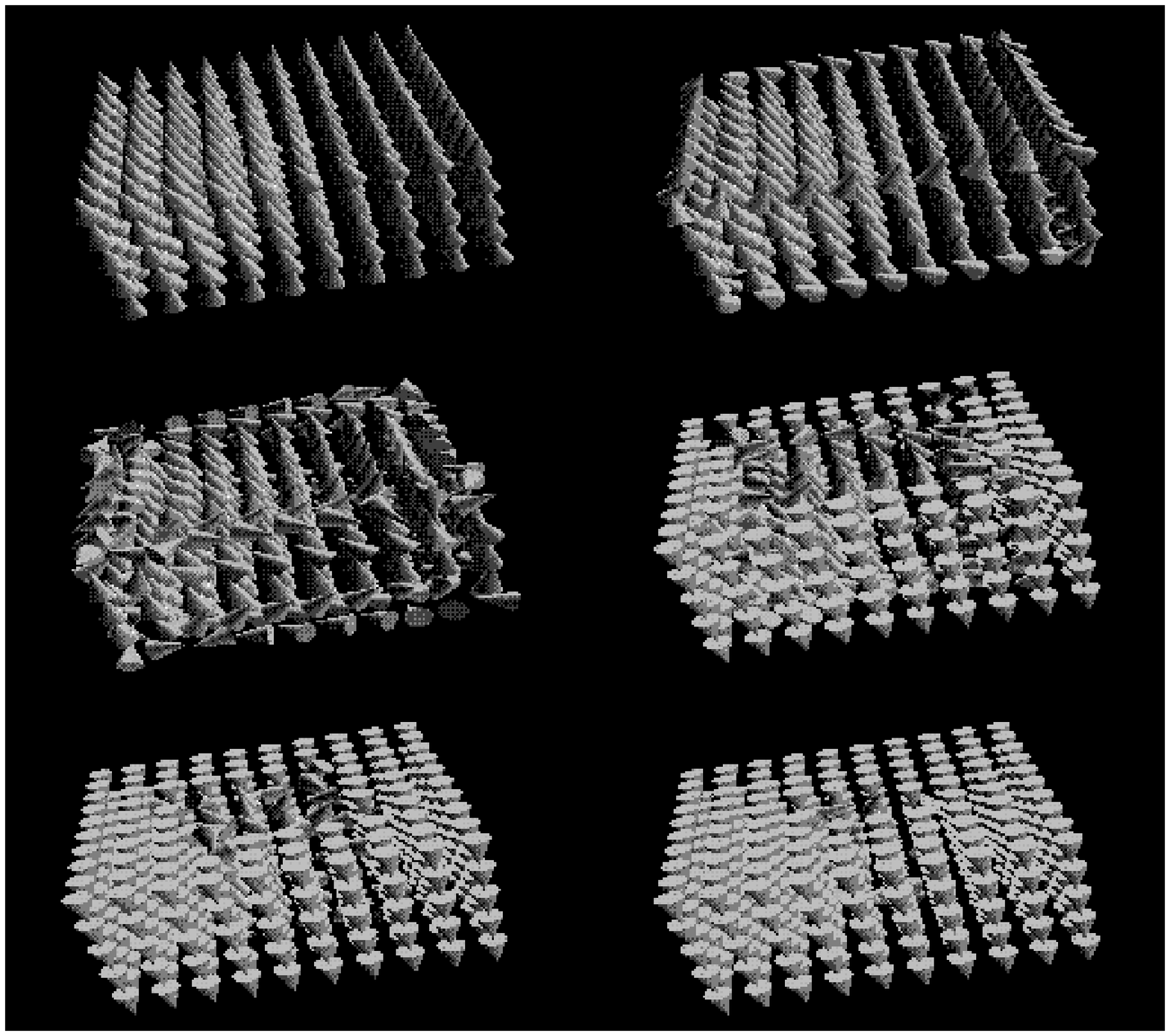}
\caption[]{
A series of snapshots of a
Langevin micromagnetic calculation for magnetization
reversal in a square array of
Ni pillars that are 200~nm tall, 200~nm apart, and each have a
diameter of 40~nm.  
Each pillar is discretized using 5 lattice points.
(For clarity of presentation, the vertical scale of this figure
is enhanced compared with the horizontal scale.)
The temperature is 300~K, the spins are initially
up, and the
applied field is down with a magnitude of $1225$~Oe.
The integration time step is $\Delta t$$=$1~psec.
The time sequence following field reversal, 
(reading line by line -- left to right and top to bottom) 
is 4~nsec, 8~nsec, 12~nsec, 16~nsec, 20~nsec, and 24~nsec.  
This reversal mode is different from  
coherent rotation [\MALL , \NEEL ].  
}
\end{figure}

\vskip 15 truept
\noindent {\bf CONCLUSIONS}

Realistic simulations of models for 
magnetic materials have been carried out.  
The form for the probability of not switching, $P_{\rm not}(t)$, 
is shown to fit the form of a complementary error function.  
The reversal mode for arrays of single-domain magnetic pillars 
has been identified as reversal of pillars on the boundary, 
followed by reversal of interior pillars.  

\vskip 15 truept
\noindent {\bf ACKNOWLEDGMENTS}

This research was supported in part by NSF Grant No.\ 9520325, 
FSU MARTECH, and FSU SCRI (DOE Contract No.\ DE-FC05-85ER25000).  
Supercomputer access was provided by the DOE at NERSC.  

\vspace*{0.4truecm}
\noindent {\bf REFERENCES}
\vspace{0.4truecm}
\begin{enumerate}

\item[\KENT .] A.~D.\ Kent, S.\ von Moln{\'a}r, S.\ Gider, and 
D.~D.\ Awschalom,
J.\ Appl.\ Phys.\ {\bf 76},  6656  (1994).  

\item[\WIRT .] S.~Wirth, M.~Field, D.~D.\ Awschalom, 
and S.\ von~Moln{\'a}r, 
Phys.\ Rev.\ B, Rapid Communications, in press (1998).

\item[\WERN .] W.\ Wernsdorfer et al., 
Phys.\ Rev.\ Lett.\ {\bf 77}, 1873 (1996); 
Phys.\ Rev.\ B {\bf 55}, 11552 (1997); 
Phys.\ Rev.\ Lett.\ {\bf 78}, 1791 (1997).  

\item[\LEDE .] M.~Lederman, S.\ Schultz, M.\ Ozaki, 
Phys.\ Rev.\ Lett.\ {\bf 73}, 1986 (1994).  

\item[\MALL .] J.~C.\ Mallinson, 
\underline{The Foundations of Magnetic Recording} (Academic, 
New York, 1993), Second Edition.

\item[\RIKV .] P.~A.\ Rikvold, H.\ Tomita, S.\ Miyashita and 
S.~W.\ Sides,
Phys.\ Rev.\ E {\bf  49}, 5080 (1994).

\item[\RICHm .] H.~L.~Richards, et al., 
J.\ Magn.\ Magn.\ Mater.\ {\bf 150}, 37 (1995).  

\item[\RICHd .] H.~L.~Richards, et al., 
Phys.\ Rev.\ B {\bf 54}, 4113 (1996).  

\item[\RICHb .] H.~L.~Richards, et al., 
Phys.\ Rev.\ B {\bf 55}, 11521 (1997).  

\item[\RIKVr .] For a review see 
P.~A.\ Rikvold, M.~A.\ Novotny, M.\ Kolesik, and H.~L.\ Richards,
in 
\underline{Dynamical Properties of Unconventional Magnetic Systems\/}, 
edited by A.~T.\ Skjeltorp and D.~Sherrington,
NATO Science Series E: Applied Sciences, Vol.\ 349
(Kluwer, Dordrecht, 1998).

\item[\NOVO .] M.~A.\ Novotny, Phys.\ Rev.\ Lett.\ {\bf 74}, 
1 (1995), Erratum {\bf 75}, 1424 (1995); 
M.\ Kolesik, M.~A.\ Novotny, and P.~A.\ Rikvold, 
Phys.\ Rev.\ Lett.\ {\bf 80}, 3384 (1998).  

\item[\VOTE .] A.~F.\ Voter, Phys.\ Rev.\ Lett.\ {\bf 78},  3908  (1997); 
J.\ Chem.\ Phys.\ {\bf 106},  4665  (1997).

\item[\AHAR .] A.\ Aharoni, 
\underline{Introduction to the Theory of Ferromagnetism}, 
(Clarendon Press,\\ Oxford, 1996).  

\item[\BROWWF .] W.~F.\ Brown, Phys.\ Rev.\ {\bf 130}, 1677  (1963).

\item[\BIND .] K.~Binder, in 
\underline{Monte Carlo Methods in Statistical Physics}, 
edited by K.~Binder,\\ (Springer, Berlin, 1979).

\item[\MART .] P.~A.\ Martin, J.\ Stat.\ Phys.\ {\bf 16}, 149 (1977).  

\item[\BOER .] E.~D.\ Boerner and H.~N.\ Bertram,
IEEE Trans.\ Magn.\ {\bf 33}, 3052 (1997).

\item[\GARA .] D.~A.\ Garanin, Phys.\ Rev.\ B {\bf 55}, 3050 (1997).

\item[\BROWG .] G.\ Brown, M.A.\ Novotny, and P.A.\ Rikvold, in preparation.

\item[\NEEL .] L.~N\'eel, Ann.\ Phys., Paris {\bf 3}, 137 (1948).

\end{enumerate}


\end{document}